\documentclass[pra,aps,reprint,showpacs,amsmath,superscriptaddress]{revtex4-1}

\usepackage{bm}
\usepackage{graphicx}
\usepackage{dsfont}

\usepackage{bm}
\usepackage{mathrsfs}

\usepackage{framed}

\usepackage[all]{xy}
\usepackage{graphicx}
\usepackage{framed}
\usepackage{comment}
\usepackage{here}                    
\usepackage{latexsym}		     
\usepackage{amsmath}     
\usepackage{amsfonts}  
\usepackage{amssymb}
\usepackage{amsthm}
\usepackage{dcolumn}
\usepackage{color}
\usepackage{mathrsfs}

\renewcommand{\>}{\rangle}
\newcommand{\<}{\langle}

\newcommand{\ket}[1]{|#1\>}
\newcommand{\bra}[1]{\<#1|}

\newcommand{\be}{\begin{equation}}
\newcommand{\ee}{\end{equation}}
\newcommand{\bea}{\begin{eqnarray}}
\newcommand{\eea}{\end{eqnarray}}

\newcommand{\ha}{\hat{a}}

\usepackage[colorlinks,citecolor=blue,linkcolor=blue,urlcolor=blue]{hyperref}

\begin{document}

\title{Photon-assisted quantum state transfer and entanglement generation in spin chains}

\author{A. Gratsea}
\affiliation{Department of Physics, University of Crete, P.O. Box 2208, GR-71003 Heraklion, Crete, Greece}

\author{G. M. Nikolopoulos}
\email{nikolg@iesl.forth.gr}

\affiliation{Institute of Electronic Structure \& Laser, FORTH, P.O. Box 1385, GR-70013 Heraklion, Greece}

\author{P. Lambropoulos}
\affiliation{Department of Physics, University of Crete, P.O. Box 2208, GR-71003 Heraklion, Crete, Greece}
\affiliation{Institute of Electronic Structure \& Laser, FORTH, P.O. Box 1385, GR-70013 Heraklion, Greece}

\date{\today}

\begin{abstract}
We propose a protocol for state transfer and entanglement generation between two distant spin qubits (sender and receiver) that have different energies. The two qubits are permanently coupled to a far off-resonant spin-chain, and the qubit of the sender is driven by an external field, which provides the energy required to bridge the energy gap between the  sender and the receiver. State transfer and  entanglement generation are  achieved via virtual single-photon and multi-photon transitions to the eigenmodes of the channel. 
\end{abstract}

\pacs{
03.67.Hk,
03.67.Lx,
42.50.−p
}

\maketitle


\section{INTRODUCTION} 
\label{sec1}

Faithful transfer and distribution of quantum states between two distant sites are of 
vital importance for various architectures of quantum information processing (QIP), 
which rely on the interconnection of different processors and memories.  
The use of photons as mediators is not an easy task for  
QIP schemes with matter-based  ``quantum hardware", such as 
trapped ions or atoms, spins in solid-state systems, electrons in quantum dots,  
Josephson junctions, etc \cite{QCQI}. 
In this respect, qubit chains have been proposed as quantum channels and entanglers \cite{BosePRL03}, 
so that information can be stored, transferred and processed utilizing the same quantum  
hardware, thereby avoiding the need for reliable interfaces between 
photonic and matter qubits. 

The typical scenario is usually formulated in terms of spin qubits, and 
pertains to two distant terminal qubits, which are connected 
to the ends of a spin chain that acts as a quantum channel. 
The main task is then to find Hamiltonians, which ensure the faithful 
transfer of an arbitrary qubit state or the entanglement generation 
between the two terminal qubits, at a well-prescribed time. 
Such types of problems have attracted considerable attention 
over the last decade or so, and an abundance of protocols have been proposed \cite{review1,review2,review3,OurBook}. 
Depending on whether external control is required for the transfer of the state (besides the initialization and the read-out processes which are always present), one can distinguish between passive and active schemes. Passive schemes do not require any external control 
to perform their task, and they typically involve a judicious engineering of the 
couplings in the entire system (terminal qubits+channel)  \cite{NikJPCM04,Christandl,stolze1,Shi05,JungPRA05,CapPRA11}. 
The engineering of the couplings aims at a commensurate linear or 
quadratic spectrum, which in view of the centrosymmetry of the system ensures the 
ideally perfect transfer of states between the terminal qubits. In general, however, faithful (not perfect) state transfer, does not require such an extensive engineering, 
and can be achieved by adjasting only the couplings of the terminal qubits to the channel  \cite{Apollaro,weakRef1,HarReuPle06,YaoPRL11}. 
Active state-transfer schemes rely on a  judicious sequence of operations and/or measurements that are applied individually or collectively on the qubits of the system  \cite{PetLamOC06,DiFraPatKimPRL08,adiabatic,ZwiAlvBenKurNJP14,BurBos05,BurGioBos07,SchPem09,LorPRA13}. 
The same principles and techniques can be exploited for the generation of entanglement 
between the terminal qubits \cite{JungPRA05,HarReuPle06,NikJPCM04b, Banchi, VenPRA07,VenPRL06,EstPRA}.

The majority of the proposed protocols for state transfer and 
entanglement generation pertains to resonant terminal qubits (i.e., for terminal qubits that have the same energy), which from a practical point of view, requires 
a simultaneous control of both qubits. Small deviations from this rule have been considered only in the 
context of weak imperfections (i.e., static diagonal disorder) that perturb the otherwise ideal 
performance of the protocols (e.g., see \cite{EstPRA,dis1,dis2,dis3}). 
None of the aforementioned protocols, however, is designed to  
work for detuned terminal qubits, that is for terminal qubits that have different energy. 
Hence, their performance deteriorates  
as the  energy difference between the terminal qubits becomes comparable to or larger than the couplings to the channel \cite{dis1,dis2,dis3}. 
To the best of our knowledge, for the time being,  state transfer and entanglement generation 
between two  terminal qubits of different energies, remain open questions.

\begin{figure*}
\includegraphics[scale=0.45]{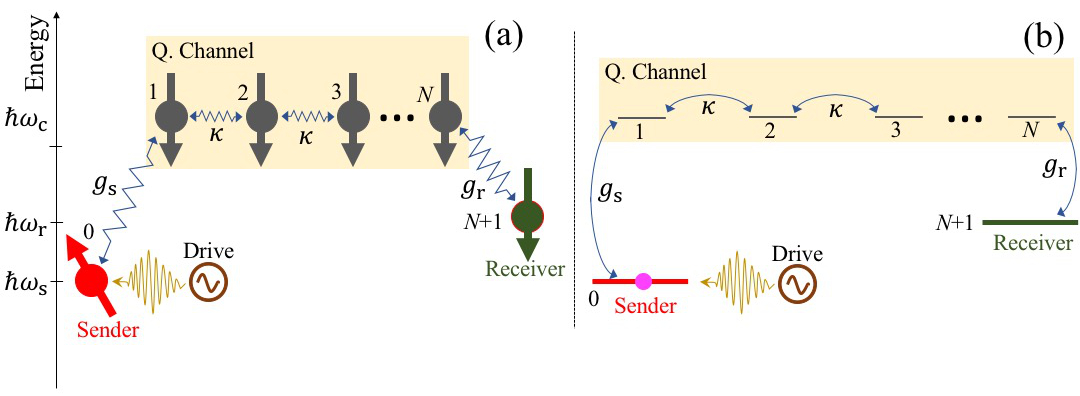}
\caption{(Color online)  Schemetic representation of the two equivalent systems under consideration. (a) 
Two distant detuned spin qubits, the sender and the receiver, are permanently coupled to a far-off resonant spin chain, which is initially prepared in its ground (vacuum) state. The qubit of the sender is initially prepared in a qubit state $\ket{\psi}$ and it is driven by an external field. (b) Employing the Jordan-Wigner transformation, the system pertains to the transfer of a single excitation between two detuned discrete states, via a far-off resonant array of $N$ coupled empty states. Note the energy difference between the sender and the receiver (i.e., $\omega_{\rm s}\neq \omega_{\rm r}$) for both (a) and (b).}
\label{fig1}
\end{figure*}

Our aim in this work is to propose a  faithful state-transfer scheme, which provides an answer to this question. The proposed scheme pertains to detuned terminal qubits (i.e., terminal qubits with different energies/frequencies), which are permanently coupled to a far off-resonant qubit-chain channel. The qubit of the sender is driven by an external field, which allows the transfer of the state from the sender  to the receiver at a prescribed time, when the photon energy (or multiple of it) matches the corresponding energy difference (hence the term photon-assisted transfer). The transfer is mediated by virtual excitation of the channel's modes. Tuning the amplitude of the driving and the photon frequency one can control the number of photons involved in the process, while  by adjusting  the duration of the driving one can achieve entanglement generation between the terminal qubits. Although a single-photon process, if possible, would be preferable in many cases,    
our results provide a solution for any number of photons, and the pertinent conditions are discussed. 

From a formal point of view, this tunability stems from the fact that the driving essentially renormalizes the coupling of the sender's qubit to the first qubit of the channel. In the past, 
such dynamical renormalization has been investigated in the contexts of coherent 
destruction of tunneling \cite{CDT1,CDT2}, and photon-assisted mesoscopic transport \cite{QD1,QD2}. Moreover, there have been studies on the fast preparation of quantum states \cite{WubsCP10}, and the control of decoherence  \cite{ChongPRA15}. 
 In the framework of QIP, the coupling renormalization has been exploited for 
temporal controlled suppression of couplings between adjacent qubits, so as  
to achieve directional transfer of quantum states and the generation of entanglement in qubit chains 
\cite{CrePRL07,ZuecoPRA09}.  By contrast, 
here we  demonstrate how to use the dynamical renormalization not for suppression, but rather for activation of an indirect coupling between the spatially separated terminal qubits, 
thereby facilitating the faithful transfer of quantum information between them, which would not have been possible by means of other known protocols. 

Furthermore, due to the presence of external driving the problem of 
state transfer acquires new aspects pertinent to the validity of the rotating-wave approximation (RWA) \cite{ShoreBook,LPbook,HuangPRA17}. 
This approximation has not been of relevance for the state-transfer protocols that have been discussed in the literature so far, but it plays a central role in the protocol under consideration. The present theoretical framework is rather general and goes beyond the 
RWA. Hence, it allows us to investigate the conditions under which such an approximation 
can be applied in the context of the system under consideration. These conditions are of 
particular importance for a potential realization of the proposed protocol in various physical platforms.

The paper is organized as follows. 
After a brief description of the system, Sect.  \ref{sec2} provides the detailed  formal framework for the problem under consideration. Section \ref{sec3} contains the bulk of the numerical results, with the detailed discussion of the 
interplay of the various parameters that determine the evolution  of the system. A summary with concluding remarks is given in Sec. \ref{sec4}.


\section{THEORETICAL FRAMEWORK}
\label{sec2}

Given that various physical realizations of qubits are currently pursued world-wide \cite{QCQI}, our results will be presented in a generic theoretical framework. Our qubits represent generic two-level systems, and 
for the convenience of exposition we will also use the term spin as it provides a simple physical picture of the network. 
The system under consideration is depicted in Fig. \ref{fig1}(a). Two distant qubits labelled by 0 and 
$N+1$ (to be referred to hereafter as the sender and the receiver),  are permanently coupled to 
a quantum channel consisting of a  chain of $N$ permanently coupled spins, while the 
qubit of the sender is driven by an external field.  The evolution of the channel is governed by 
an XX spin-chain Hamiltonian, 
and by applying the Jordan-Wigner transformation, the problem can be formulated in terms of 
free spinless fermions \cite{OurBook,review3,stolze1}. In this picture, the ``down" and ``up" states of 
a spin are viewed as empty and singly occupied fermion states i.e., 
\bea
\ket{\downarrow}\equiv\ket{0},\,\,\ket{\uparrow}\equiv\ket{1}. 
\label{JWpicture}
\eea
Hence, the generic 
basis states for a qubit are $\lbrace \vert 0\rangle,\vert 1\rangle\rbrace$, and the equivalent 
system is depicted in Fig. \ref{fig1}(b). 

\subsection{Hamiltonian}
\label{sec2a}

The full Hamiltonian of the system is 
\bea
{\cal H}  = {\cal H}_{\rm 0}+{\cal H}_{\rm d}+{\cal H}_{\rm c} +{\cal V}, 
\eea
where ${\cal H}_{\rm 0}$ is the unperturbed Hamiltonian of the sender and the receiver, 
${\cal H}_{\rm c}$ refers to the channel, ${\cal V}$ describes 
the interaction between the sender/receiver qubits and the channel, while ${\cal H}_{\rm d}$ 
describes the driving of the qubit of the sender. More precisely, we have 
\begin{subequations}
\bea
\label{HamH0} 
{\cal H}_{\rm 0} &=& \hbar\omega_{\rm s} \hat{a}^{\dag}_{\rm s} \hat{a}_{\rm s} +  
 \hbar\omega_{\rm r} \hat{a}^{\dag}_{\rm r} \hat{a}_{\rm r},  
 \\
{\cal H}_c  &=&  \sum_{i=1}^{N} \hbar\omega_{\rm c} \hat{a}^{\dag}_i \hat{a}_{i} 
 + \sum_{i=1}^{N-1} \hbar \kappa ( \ha^{\dag}_i \ha_{i+1} + \ha^{\dag}_{i+1} \ha_i), 
\label{HamHub} 
\\
{\cal V}  &=&  \hbar g_{{\rm s}} 
(\ha^{\dag}_{\rm s} \ha_{1} + \ha^{\dag}_{1} \ha_{\rm s})+\hbar g_{{\rm r}} 
( \ha^{\dag}_{N} \ha_{\rm r}+\ha^{\dag}_{\rm r} \ha_{N}),
\label{HamV:eq} 
\\
{\cal H}_{\rm d} &=&  \frac{V_{\rm d}}{2} f(t) \cos(\omega t) \hat{a}_{\rm s}^\dag \hat{a}_{\rm s}.
\label{Hd:eq}
\eea
\end{subequations}
The creation (annihilation) operator $\ha^{\dag}_i (\ha_i)$ creates (annihilates) an excitation at the $i$th site of the channel, with energy $\hbar\omega_{\rm c}$. 
The corresponding operators for the sender and the receiver are denoted by 
$\ha^{\dag}_{\rm s} (\ha_{\rm s})$ and $\ha^{\dag}_{\rm r} (\ha_{\rm r})$, and the energies are $\hbar\omega_{\rm s}$ and  $\hbar\omega_{\rm r}$, respectively. Throughout this work we are interested in 
detuned terminal qubits, with the corresponding energy difference denoted by 
$\omega_{\rm r,s}:=\omega_{\rm r}- \omega_{\rm s}$.  The coupling between adjacent sites in the chain is denoted by $\kappa$, while the sender and the receiver are coupled to the two outermost 
sites with couplings $g_{{\rm s}}$  and $g_{{\rm r}}$, respectively. We assume a pulsed driving, with amplitude $V_{\rm d}$, while $f(t)$ is the pulse shape, and $\omega$  the frequency of the driving field. 

The driving term can be treated through the unitary transformation 
${\widetilde{\cal H}}(t) = 
{\cal W}(t)\left ({\cal H}(t)-i\hbar\partial_t \right ){{\cal W}}^{\dagger}(t) 
$ \cite{QD2,ZuecoPRA09,APbook}, with 
${\cal W}(t) := \exp \left[  {\rm i} h(t) \hat{a}_{\rm s}^\dag \hat{a}_{\rm s}  \right ]$ and  
\be
h(t) := \frac{V_d}{2\hbar} \int_0^{t} f(t^\prime)\cos(\omega t^\prime) 
\mathrm{d}t^\prime.
\label{ht:eq}
\ee
The new Hamiltonian is then  
${\widetilde{\cal H}} ={\cal H}_0+{\cal H}_{\rm c} + {\widetilde{\cal V}}$, 
where
\begin{subequations} 
\label{HamV2:eq} 
\bea
{\widetilde{\cal V}}  &=&  \hbar 
(\tilde{g}_{{\rm s}}  \ha^{\dag}_{\rm s} \ha_{1} + \tilde{g}_{{\rm r}}  \ha^{\dag}_{N} \ha_{\rm r}+{\rm H.c.}),
\eea
with  
\be
\tilde{g}_{{\rm s}} := g_{{\rm s}} e^{{\rm i}h(t)},\,\, {\rm and}\,\,\, \tilde{g}_{{\rm r}} := g_{{\rm r}}. 
\label{tilde_gs}
\ee
\end{subequations}

In practise we are interested in smooth pulses of some duration $\tau$,  
that rise and and drop slowly. Hence, the pulse shape satisfies 
 $f(0) =  0$, $f(\infty) = 0$ (where the limit $t\to\infty$ is attained for times $t\gg \tau$), as well as 
 the adiabaticity condition  
\bea
\left | \frac{{\rm d}f}{{\rm d}t} \right | \ll \omega |f(t)|.  
\eea 
Under these conditions, one readily  obtains a rather simple form for the function $h(t)$ that enters 
Eqs. (\ref{HamV2:eq}) namely 
\bea
h(t) \simeq\frac{V_{\rm d}}{2\hbar\omega } f(t) \sin(\omega t) := z(t) \sin(\omega t). 
\label{ht2:eq}
\eea
In the following analysis the profile of the pulse is chosen so that its maximum value is 1. Hence, the maximum amplitude of the drive is given by $z_0:=V_{\rm d}/(2\hbar\omega)$.

\subsection{Equations of motion}
\label{sec2b}

At time $t=0$ the qubit of the sender is prepared in the state 
\bea
\ket{\psi}_{\rm s} = \alpha\ket{0}_{\rm s}+\beta\ket{1}_{\rm s},
\eea
which has to be transferred to the qubit of the receiver after a prescribed time, say  $T>\tau$. 
The  channel is initially prepared in its ground state 
with zero excitations $\ket{{\bf 0}}_{\rm c}:=\ket{0_1,\ldots,0_N}$, while the qubit of the receiver is in the state $\ket{0}_{\rm r}$. 
Hence, the total initial state of the system is 
$
\ket{\Psi(0)} = \ket{\psi}_{\rm s}\otimes\ket{{\bf 0}}_{\rm c}\otimes \ket{0}_{\rm r}.
$ 
For later convenience, 
we   introduce the zero-excitation state of the entire system 
$\ket{{\bf 0}}:=\ket{0}_{\rm s}\otimes\ket{{\bf 0}}_{\rm c}\otimes\ket{0}_{\rm r}$. 
The single-excitation subspace of the Hilbert space is spanned by the states 
$\{\ket{{\bf 1}}_j \}$, where 
\[\ket{{\bf 1}}_j := \ha^{\dag}_j\ket{\bm{0}},\quad {\rm for}\quad j=0, 1, \ldots, N+1.\]
Based on the aforementioned convention, the states of the sender and 
the receiver correspond to $j=0$ and $j=N+1$, respectively. 
Hence, the initial state of the system can be expressed as
\bea
\ket{\Psi(0)} = \alpha \ket{\bm{0}} + \beta \ket{{\bf 1}}_{\rm s},
\eea 
where by definition  $\ket{{\bf 1}}_{\rm s} := \ha^{\dag}_{\rm s}\ket{\bm{0}}$ i.e., 
the subscript  ${\rm s}$ denotes the position of the excitation.

Since the Hamiltonian preserves 
the number of excitations, we need to consider 
only the zero- and single- excitation subspaces of the total Hilbert space.
Then the system evolves in time as 
$\ket{\Psi(t)} = \widehat{\cal U}(t)  \ket{\Psi(0)} = \alpha \ket{\bm{0}} 
+ \beta \sum_{j} A_j(t) \ket{{\bf 1}}_j$, 
where 
\[ \widehat{\cal U}(t) = \mathcal{T} \exp \left [\frac{1}{i \hbar} \int_{0}^t \!\!  
\widehat{\widetilde {\cal H}}(t') dt' \right ]
\]
 is the (time-ordered, $\mathcal{T}$) evolution operator.
Clearly, only the states in the single 
excitation sub-space $\{\ket{{\bf 1}}_j\}$ evolve in time with the 
corresponding amplitudes $A_j(t) \equiv ~_j\bra{{\bf 1}} \widehat{\cal U}(t)  \ket{{\bf 1}}_{\rm s}$,
while the vacuum (or ground) state $\ket{\bm{0}}$ remains unchanged. 
The amplitudes $A_0$ and $A_{N+1}$ refer to the qubits of the sender and the receiver, 
while the amplitudes $A_j$ with $1\leq j\leq N$, to the channel. 

In the interaction picture \cite{LPbook}, the equations of motion for the amplitudes read
\begin{subequations}
\label{em1e_v1}
\be
{\rm i} \frac{{\rm d} A_j}{{\rm d}t} = 
G_{j-1}^\star e^{{\rm i}(\omega_j - \omega_{j-1}) t} A_{j-1} + G_{j} e^{{\rm i}(\omega_j - \omega_{j+1}) t}  A_{j+1} , 
\ee 
where 
$\omega_0 = \omega_{\rm s}$, $\omega_{N+1}= \omega_{\rm r}$, 
$\omega_{j}= \omega_{\rm c}$ for all $1\leq j\leq N$, and the couplings are given by 
\begin{equation}
G_{l} = \left\{
\begin{array}{lr}
g_{\rm s} e^{{\rm i}h(t)} & \text{if } l = 0,\\
g_{\rm r}  & \text{if } l = N,\\
\kappa  & \text{if } 1\leq l\leq N-1,\\
0  & \textrm{otherwise}.
\end{array} \right.
\label{coup1:eq}
\end{equation}
\end{subequations}

The set of Eqs. (\ref{em1e_v1}), together with Eq.  (\ref{ht2:eq}) fully describe the evolution of the system for smooth pulses,  
and can be solved numerically using standard techniques. 
However, further insight into the physics 
of the system can be obtained by applying the 
Jacobi-Anger expansion so that to rewrite the coupling for the 
driven qubit as
\be 
\label{eq:6}
G_{\rm 0} = \tilde{g}_{\rm s} = g_{\rm s} \sum_{\substack{n={-\infty}}}^\infty {\cal J}_n ( z ) e^{i n \omega t}:=  \sum_{\substack{n={-\infty}}}^\infty g_{\rm s}^{(n)} e^{i n \omega t},
\ee
where ${\cal J}_{n} $ is the $n$th order Bessel functions of the first kind, and $z = z_0 f(t)$. 
This expansion shows that no RWA has been applied in Eqs. (\ref{em1e_v1}), and  all possible $n-$photon transitions to the channel contribute in the transfer of the state from the sender to the receiver. 

\begin{figure}
\includegraphics[scale=0.35]{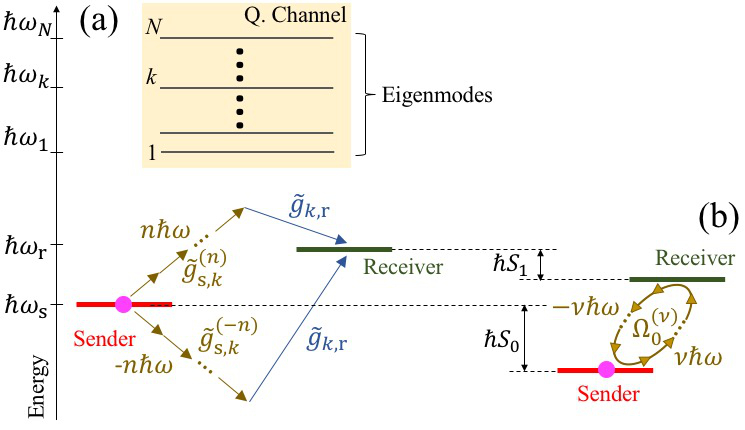}
\caption{(Color online)  Schematic representation of the system under consideration. (a)  After 
diagonalization of the Hamiltonian of the channel, the system of Fig. \ref{fig1}(b) 
reduces to the transfer of an excitation between to detuned levels, 
via virtual excitation of a far off-resonant band of empty states, pertainng to the eigenmodes of the channel. 
In general, the coupling of the sender to the $k$th eigenmode involves absorption or emission of 
$ n$ photons, for all possible $n=0,1,2,\ldots$. Adjusting the strength of the driving $z_0$ and the 
photon frequency $\omega$, one can control the $n-$photon transition(s) that dominate the dynamics. 
(b)  Eliminating the modes of the channel adiabatically, one obtains an effective two-level system pertaining 
to the shifted states of the sender and the receiver. The standard $\nu-$photon Rabi model is obtained 
when the RWA is valid. }
\label{fig2}
\end{figure}

This point becomes clearer if one adopts an alternative, yet fully equivalent, point of view. 
Diagonalizing the Hamiltonian for the channel one obtains a new 
Hamiltonian, which describes the transfer of the state from the sender to the receiver, 
through a band of states pertaining to the eigenmodes of the channel (see appendix \ref{app1}).  
A schematic representation of the system in the new picture is given in Fig. \ref{fig2}.

The corresponding equations  of motion are given by 
\begin{subequations}
\label{em1e_v2}
\bea
\label{em1e_v2:eq1}
{\rm i} \frac{{\rm d} A_{\rm s}}{{\rm d}t} &=& \sum_{k=1}^N  \sum_{n=-\infty}^{\infty} \tilde{g}_{{\rm s},k}^{(n)} (t)  e^{-{\rm i}(\omega_{k,{\rm s}}-n\omega) t}  A_{k}, \\
{\rm i} \frac{{\rm d} A_{\rm r}}{{\rm d}t} &=& \sum_{k=1}^N \tilde{g}_{{\rm r},k}
e^{-{\rm i} \omega_{k,{\rm r}}t} A_{k}, 
\label{em1e_v2:eq2}\\
{\rm i} \frac{{\rm d} A_k}{{\rm d}t} &=& \sum_{n^\prime=-\infty}^{\infty} \tilde{g}_{{\rm s},k}^{(n^\prime)} 
e^{{\rm i}(\omega_{k,{\rm s}}-n^\prime\omega) t} A_{\rm s} 
+ \tilde{g}_{{\rm r},k} e^{{\rm i}\omega_{k,{\rm r}} t}  A_{\rm r},\nonumber \\
\label{em1e_v2:eq3}
\eea 
\end{subequations}
where $\omega_{k,x}:=\omega_k-\omega_x$ for $x\in\{{\rm s},{\rm r}\}$.  
The amplitude $A_k$ refers to the mode with momentum $k$, while 
$\tilde{g}_{{\rm s},k}^{(n)}$ and $\tilde{g}_{{\rm r},k}$ are the couplings of the sender and the receiver to this mode (see appendix \ref{app1} for the related expressions). It is  clear now that the transfer of the 
state is mediated by the excitation of the channel with the emission and absorption of $n$ photons [see Fig. \ref{fig2}(a)]. 
The extension of the summations from $-\infty$ to $\infty$ shows that the present formalism so far  does not involve the RWA, and includes all possible $n-$photon 
transitions, real and virtual. Which ones will turn out to be dominant is determined by 
the coupling strengths $\tilde{g}_{{\rm s(r)},k}$ and the photon energy $\hbar\omega$, 
with respect to the 
energy separation of the band from the sender and the receiver.

\begin{figure}
\includegraphics[scale=0.45]{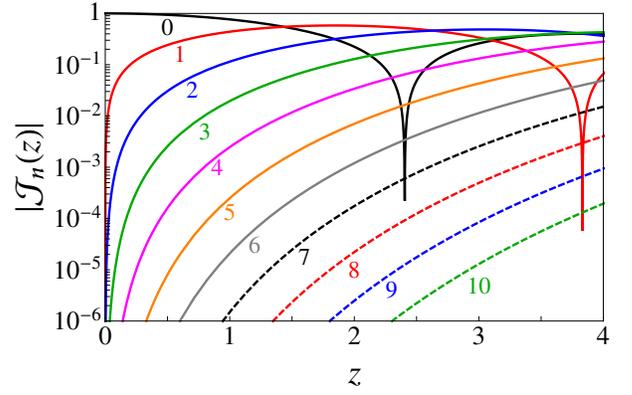}
\caption{(Color online)  Logarithmic plot of $|{\cal J}_n(z)|$ as a function of $a$, for different values of the order $z$. For Bessel functions with negative $n$ one may recall the identity 
${\cal J}_{-n}(z) = (-1)^n {\cal J}_n(z)$.}
\label{fig3}
\end{figure}

As shown in appendix \ref{app1}, for a given driving $z(t)$, 
the coupling of the sender to the $k$th eigenomode via $n-$photon absorption (or emission) 
is proportional to the $n$th order Bessel function i.e.,  $|\tilde{g}_{{\rm s},k}^{(n)}|\propto 
|{\cal J}_n(z)|$.
Throughout this work we focus on moderate to weak amplitudes of driving, corresponding to $0\leq z_0< 4$. As depicted in Fig. \ref{fig3}, $|J_n(z)|\leq 1$ for all $z\geq 0$, while for a given value of $z(t)$,  the values of the Bessel functions of different orders $n$, may differ by orders of magnitude. 
These observations suggest that, by adjusting the amplitude of the driving $z_0$, one 
can control the $n-$photon 
transition(s) that dominate the dynamics of the system within the prescribed time of the transfer.   
Effects of weaker couplings are expected to become important for 
larger time scales, and can be neglected. As will be seen below, for the values of $z_0$ under consideration,  our results are well-explained in terms of few-photon transitions (i.e., $0\leq n \leq 5$).

\subsection{Effective two-level system (TLS)}
\label{sec2c}
The proposed schemes for state transfer and entanglement generation 
rely on setting the detuned terminal qubits to be far off-resonant from the channel. In this limit,  the numerical simulations presented in the following section can be interpreted in the framework of 
an effective TLS. 

We are interested in combinations of parameters such that the following conditions are satisfied  
\begin{subequations}
\label{All_assumptions}
\bea
\label{assumption1}
&&|\tilde{g}_{{\rm r},k}|\ll |\omega_{{\rm r},k}|, 
\\
\label{assumption2}
&&|\tilde{g}_{{\rm s},k}| \ll |\omega_{{\rm s},k}-n\omega|,  
\\
&&|\min\{\omega_k\}-\omega_{\rm s}|\gg \omega,
\label{assumption3}
\\
&&g_{\rm s}|{\cal J}_{n^*}(z_0)|  T \ll 1.
\label{assumption4}
\eea 
\end{subequations}
Inequality (\ref{assumption3}) implies that there is a large energy gap between the sender and the 
band of the modes, and one needs a large number of photons (say $ n=n^*\sim 10$),  
to bridge the gap and come close the band. The corresponding coupling strengths 
$|\tilde{g}_{{\rm s},k}^{(n^*)}|\propto g_{\rm s}|{\cal J}_{n^*}(z)|$ are very weak for the 
driving amplitudes of interest. For instance, the behaviour  depicted in Fig. \ref{fig3} suggests that 
depending on $z$, the  value of  
$|{\cal J}_{10}(z)|$ can be at least four orders of magnitude smaller than most of (if not all) 
Bessel functions of order $n\leq 5$. Hence, for time scales $T$ that satisfy 
inequality (\ref{assumption4}), the dynamics of the system are expected to be dominated by transitions involving $n\leq 5$ photons. The precise number of photons for the dominant transitions depends on the amplitude of the driving and the photon frequency (this point will become clear below). In any case, however,  when condition (\ref{assumption2}) is satisfied for the dominant $n$, one may expect a negligible 
excitation of the modes of the channel during the evolution.

Under these conditions, the equations for the channel can be eliminated adiabatically,  yielding  
two equations of the form (see appendix \ref{app2}) 
\begin{subequations}
\label{em1e_v3}
\bea
\label{em1e_v3:eq1}
{\rm i}\frac{{\rm d} A_{\rm s}}{{\rm d}t} &\simeq&  -S_0(t) A_{\rm s} -\Omega_0(t) A_{\rm r},
 \\
{\rm i} \frac{{\rm d} A_{\rm r}}{{\rm d}t} &\simeq& - S_1(t) A_{\rm r} -\Omega_1(t) A_{\rm s},
\label{em1e_v3:eq2}
\eea
\end{subequations}
which describe the evolution of a TLS  (consisting of the sender and the receiver), under the action of a pulsed driving that is included in $\Omega_{0(1)}(t)$. The explicit form of $S_{0(1)}$, and $\Omega_{0(1)}$ are given in appendix (\ref{app2}).  
 When there is $\nu-$photon resonance  between 
the qubits of the sender and the receiver i.e., for 
\bea
\omega_{\rm r}\simeq \omega_{\rm s}+\nu\omega  
\label{resonance:eq}
\eea 
with $\nu\ll n^*$, one can apply the RWA to obtain the standard Rabi model with 
shifts 
 \begin{subequations}
\bea
S_{0} &\simeq&  \sum_k \frac{(g_{{\rm s},k})^2}{\omega_{k,{\rm r}}},\,\,
S_{1} \simeq \sum_k \frac{(g_{{\rm r},k})^2}{\omega_{k,{\rm r}}}, \nonumber
\eea
and Rabi frequencies 
\bea
\Omega_{0}^{(\nu)}&=&\Omega_{1}^{(\nu)}\simeq  \sum_k
\frac{\tilde{g}_{{\rm s},k}^{(\nu)} \tilde{g}_{{\rm r},k} }{\omega_{k,{\rm r}}}.
\eea
\end{subequations} 

The case of $g_{\rm s} = g_{\rm r}$ is of particular interest, because it implies $S_0 \simeq S_1$. 
In this case, for $\Omega_0^{(\nu)},\Omega_1^{(\nu)}\gg |S_0 - S_1|$ (see related discussion at the end of appendix \ref{app2}), the problem can be solved analytically obtaining  \cite{ShoreBook}
\begin{subequations}
\label{eom_tls:eqs}
\bea
A_{\rm s}(t) &\simeq& \cos[\Theta(t)]
\\
A_{\rm r}(t) &\simeq& {\rm i}\sin[\Theta(t)]
\label{Ar_rwa:eq}
\eea 
with the pulse area given by 
\bea
\label{Theta_t:eq}
\Theta(t) &:=& \Theta_0 \int_0^t d t^\prime {\cal J}_\nu[z(t^\prime)],
\\
\Theta_0 &:=&  \sum_k
\frac{(-1)^{k-1}(g_{{\rm s},k})^2}{\omega_{k,{\rm r}}},
\eea
\end{subequations}
and $g_{{\rm s},k} $ given by Eqs. (\ref{gsk:eq}) and (\ref{Lki:eq}).
Hence, the evolution of the TLS at time $t$ is fully determined by the pulse area $\Theta(t)$, 
and it is independent of the shape of the pulse. 
These equations indicate that different logical gates can be applied between the sender and the receiver, by adjusting the couplings 
and the duration of the pulse \cite{QCQI}. For instance, when the system evolves for time $T\gg \tau$ and 
$\Theta(T) = \pi/2$, then the state of the combined system at the end of the evolution is 
well approximated by 
\bea
\ket{\Psi(T)} = \ket{0}_{\rm s}\otimes\ket{{\bf 0}}_{\rm c}\otimes \ket{\psi}_{\rm r}.
\label{psi_pi2}
\eea 
In this case the input state has been transferred from the sender to the receiver. 
Similarly, if the input state is $\ket{\psi} = \ket{1}$ and 
$\Theta(T) = \pi/4$, we expect the final state to be close to the maximally entangled state
\bea
\ket{\Psi(T)} = \frac{1}{\sqrt{2}}\left (\ket{1}_{\rm s}\otimes \ket{0}_{\rm r} 
+ {\rm i}\ket{0}_{\rm s}\otimes \ket{1}_{\rm r} \right ) \otimes\ket{{\bf 0}}_{\rm c}.  
\label{psi_pi4}
\eea

\begin{figure}
\includegraphics[scale=0.6]{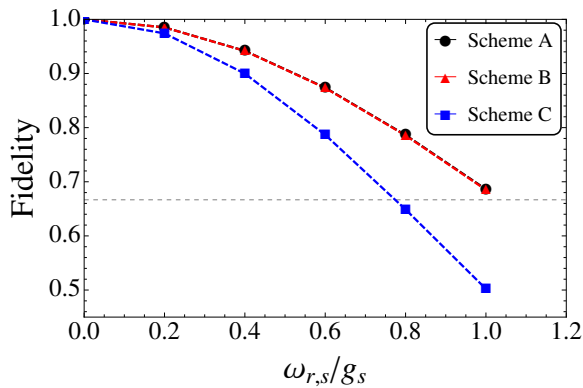}
\caption{(Color online)  Performance of three different faithful state transfer schemes for increasing detuning between the sender and the receiver, for $N = 3$. The plotted fidelity 
is ${\cal F}_{\rm min}$ (see main text).  
Scheme A refers to the protocol of Refs. \cite{NikJPCM04,Christandl}, scheme B to the protocol of Ref. \cite{Apollaro} and scheme C to the protocol of Ref. \cite{weakRef1}. In all of the cases, the sender is on resonance with the chain,  while the receiver is detuned by $\omega_{\rm r,s}$.  The horizontal dashed line marks the classical threshold of 2/3. Similar behaviour is obtained for other values of $N\geq 2$, while 
for a given value of the ratio $\omega_{\rm r,s}/g_{\rm s}$, the fidelity drops slowly with increasing $N$. }
\label{fig4}
\end{figure}

In closing this section, we would like to emphasize that the above analytic expressions for  the amplitudes and the final states are valid only for  $\Omega_0^{(\nu)},\Omega_1^{(\nu)}\gg |S_0 - S_1|$. If this condition is violated, and the effective TLS is still valid, the expressions for $A_{\rm s(r)}(t)$ 
are not given by Eqs. (\ref{eom_tls:eqs}), while their evolution may depend on the details of the pulse shape \cite{ShoreBook}. Moreover, in order to prepare the system at states (\ref{psi_pi2}) and (\ref{psi_pi4}), one has to freeze the dynamics at time $t=T$ i.e., to set $\Omega_0^{(\nu)},\Omega_1^{(\nu)}=0$. 
In view of Eqs. (\ref{gskn_tilde:eq}), (\ref{gsk:eq}) and (\ref{tilde_gs}), the coupling $\tilde{g}_{{\rm s},k}^{(\nu)}$ does not tend to 0, when $f(t)\to 0$. This is because the pulse shape $f(t)$ appears in the exponent of Eq. (\ref{tilde_gs}). Hence, the system keeps evolving after the end of the pulse, its quantum state changes, but as will be seen in the following section, the dynamics are very weak and they do not affect significantly the performance of the protocol.


\begin{figure}
\includegraphics[scale=0.35]{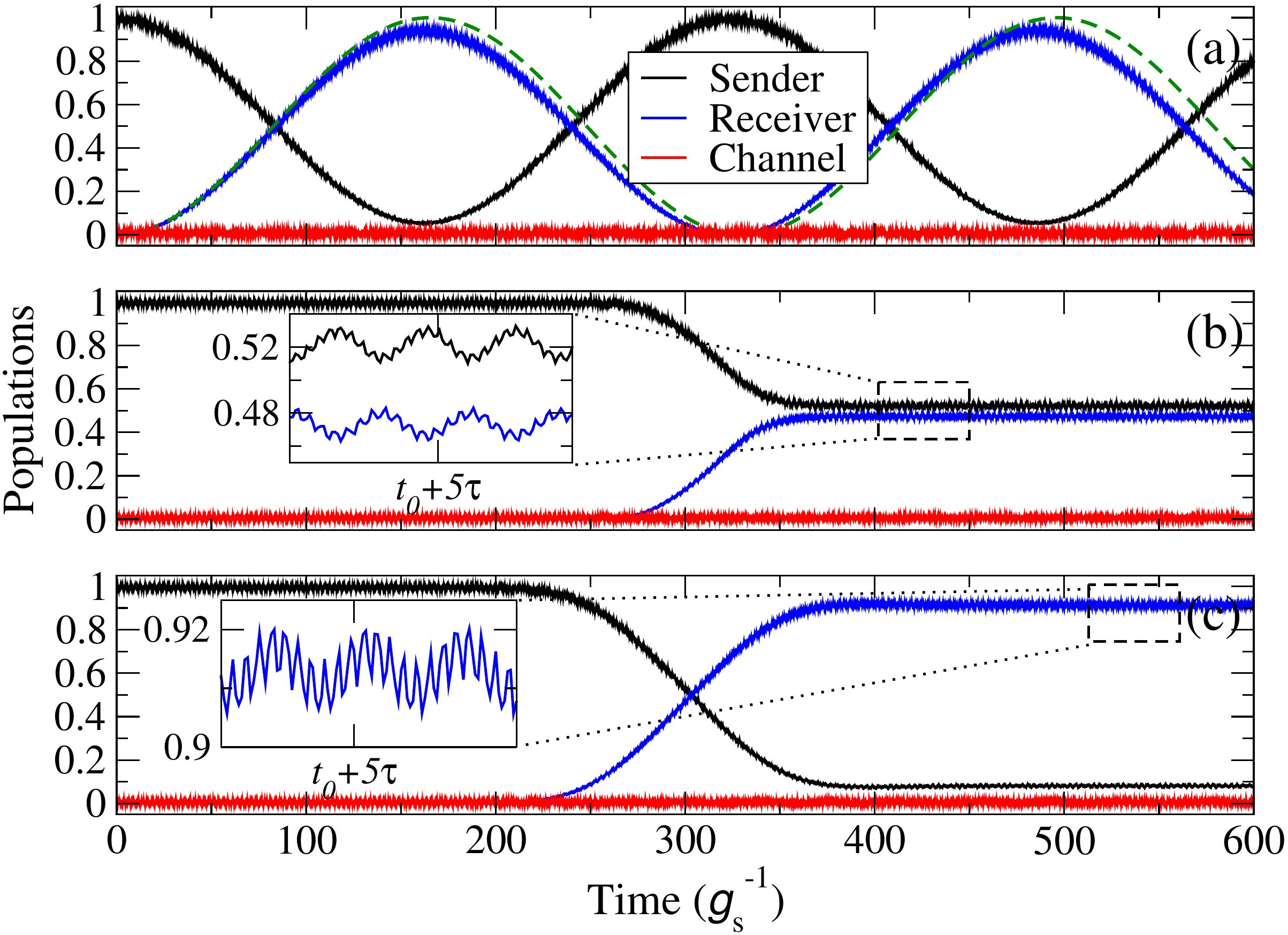}
\caption{(Color online) 
State transfer and entanglement generation mediated by a chain of $N=2$ spin qubits. 
 Numerical solution of Eqs. (\ref{em1e_v1}) for continuous $(f(t)=1)$ harmonic driving 
(a) and pulsed harmonic driving (b,c). 
Parameters of the system: $\omega_{\rm c} - \omega_{\rm s}= 22.0 g_{\rm s}$, $\omega_{\rm r} - \omega_{\rm s}=2.0g_{\rm s}$, $g_{\rm r}=1.0g_{\rm s}$, $\kappa=6.0g_{\rm s}$, $\omega = 2g_{\rm s}$ and $z_0=2.0$. Parameters of the pulse: $g_{\rm s}t_0 = 300$, and $\Theta(T)=\pi/4$;  $g_{\rm s}\tau=25.45$ (b); $\Theta(T) = \pi/2$; $g_{\rm s}\tau=50.90 $. The dashed green curve refers to Eq. (\ref{Ar_rwa:eq}). 
Time is in units of $g_s^{-1}$.  }
\label{fig5}
\end{figure}

\section{Simulations}  
\label{sec3}

The majority of the state-transfer and entanglement-generation protocols that have been discussed in the literature so far, are designed to work for resonant sender and receiver i.e., for $\omega_{\rm r,s} = 0$. As depicted in Fig. \ref{fig4}, the performance 
of three different well-studied protocols becomes worse for increasing values of  
$\omega_{\rm r,s}$ relative to $g_{\rm s}$. Moreover, we see that none of the three protocols is capable of transmitting reliably any qubit state between the sender and the receiver when  
$\omega_{\rm r,s}\gtrsim g_{\rm s}$, in the sense that the lower bound on the 
fidelity is almost equal to the classical threshold (if not smaller). The results presented in the following subsection demonstrate that the present  protocol is capable of  faithful state transfer and 
 entanglement generation between the detuned terminal qubits, 
 provided that the corresponding energy difference matches a multiple of the photon energy i.e., 
$\omega_{\rm r,s} = \nu\omega$ for $\nu>0$. In this case, the photon essentially provides the energy 
required to bridge the energy gap between the terminal qubits. 

For the sake of concreteness we have chosen a Gaussian pulse profile for the driving of the form
\bea
f(t) = \exp\left [ -\frac{(t-t_0)^2}{2\tau^2} \right ],
\eea 
where $t_0 > 5\tau $ is the center of the pulse. 
The main dynamics of the system are expected to take place  for 
times $t \in[t_0-4\tau,t_0+4\tau]$, where the main part of the Gaussian is located.

\begin{figure}
\includegraphics[scale=0.35]{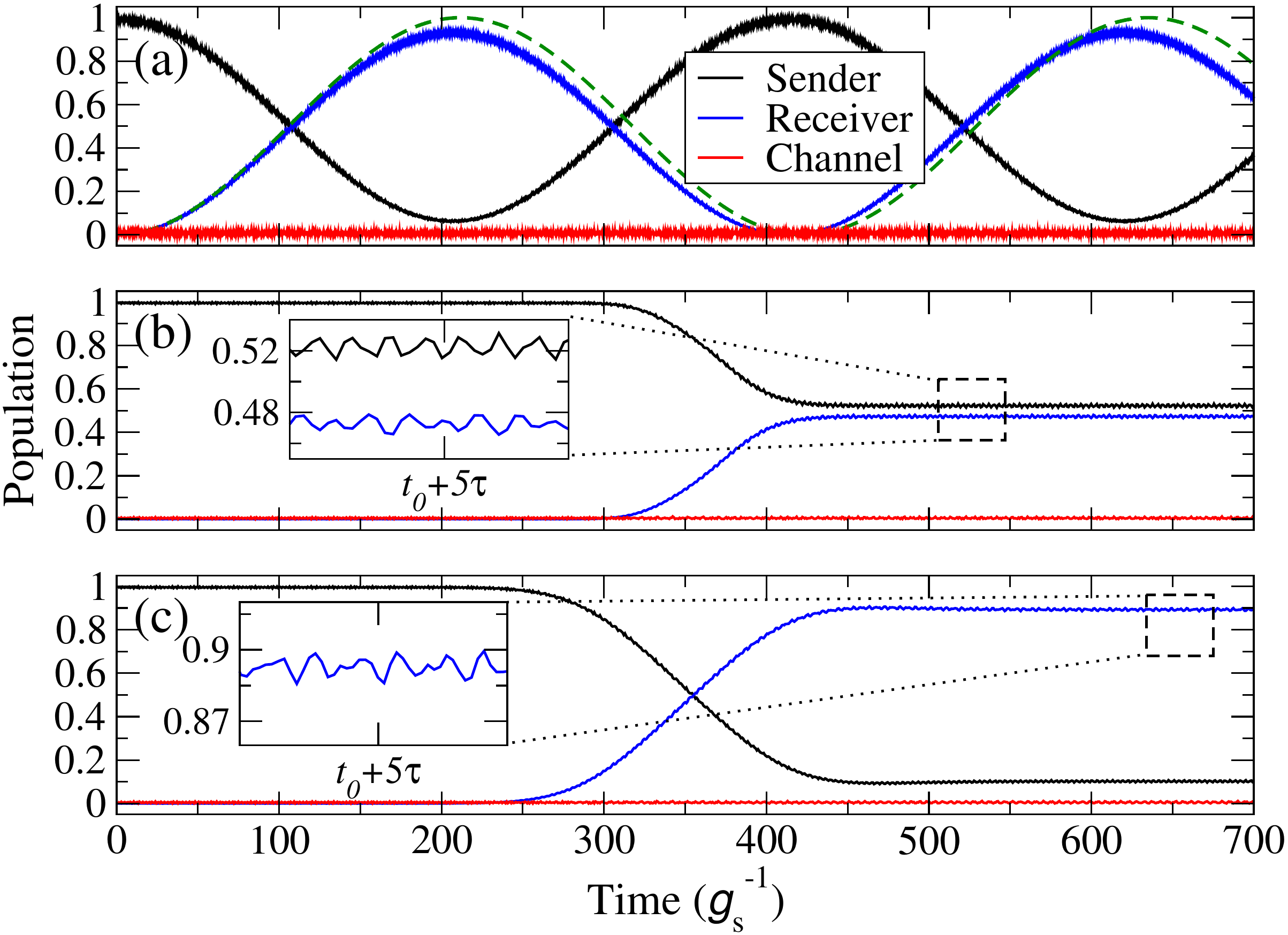}
\caption{(Color online)  State transfer and entanglement generation mediated by a chain of $N=3$ spin qubits. 
Numerical solution of Eqs. (\ref{em1e_v1}) for continuous $(f(t)=1)$ harmonic driving 
(a) and pulsed harmonic driving (b,c). 
Parameters of the system: $\omega_{\rm c} - \omega_{\rm s}= 32.0 g_{\rm s}$, $\omega_{\rm r} - \omega_{\rm s}=2.0g_{\rm s}$, $g_{\rm r}=1.0g_{\rm s}$, $\kappa=14.0g_{\rm s}$ , $\omega = 2g_{\rm s}$ and $z_0=2.0$. Parameters of the pulse: $g_{\rm s}t_0 = 350$, and $\Theta(T)=\pi/4$;  $g_{\rm s}\tau=32.66$ 
(b); $\Theta(T) = \pi/2$; $g_{\rm s}\tau=65.11$. The dashed green curve refers to Eq. (\ref{Ar_rwa:eq}). Time is in units of $g_s^{-1}$.  }
\label{fig6}
\end{figure}

\begin{figure*}
\centerline{\includegraphics[scale=0.35]{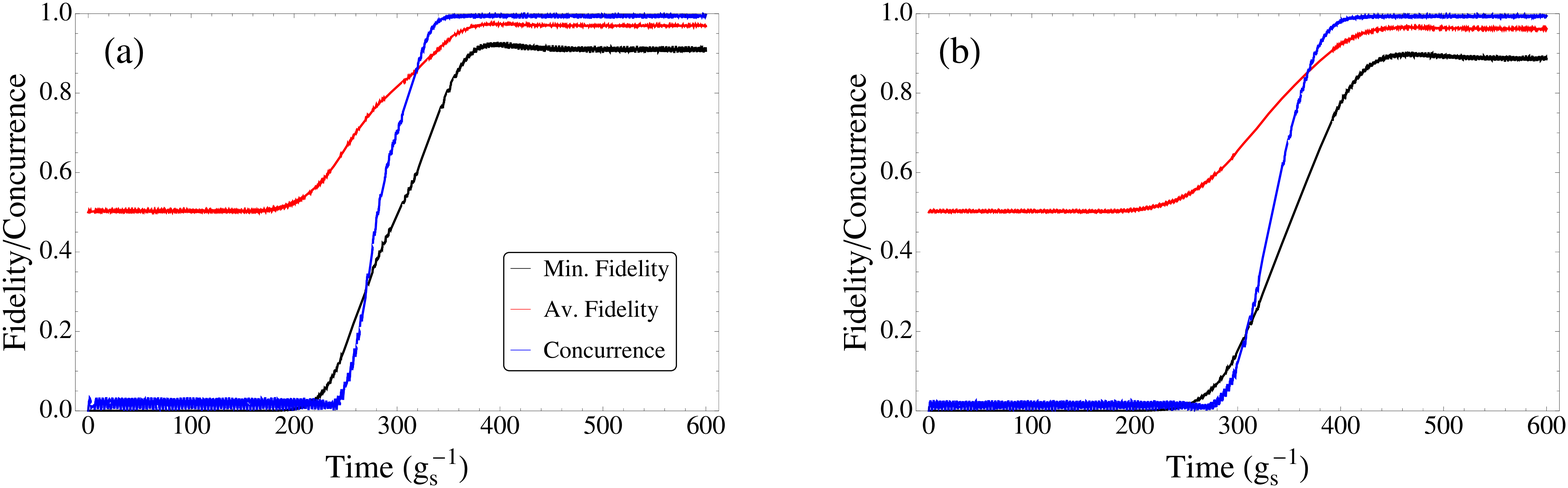}}
\centerline{\includegraphics[scale=0.35]{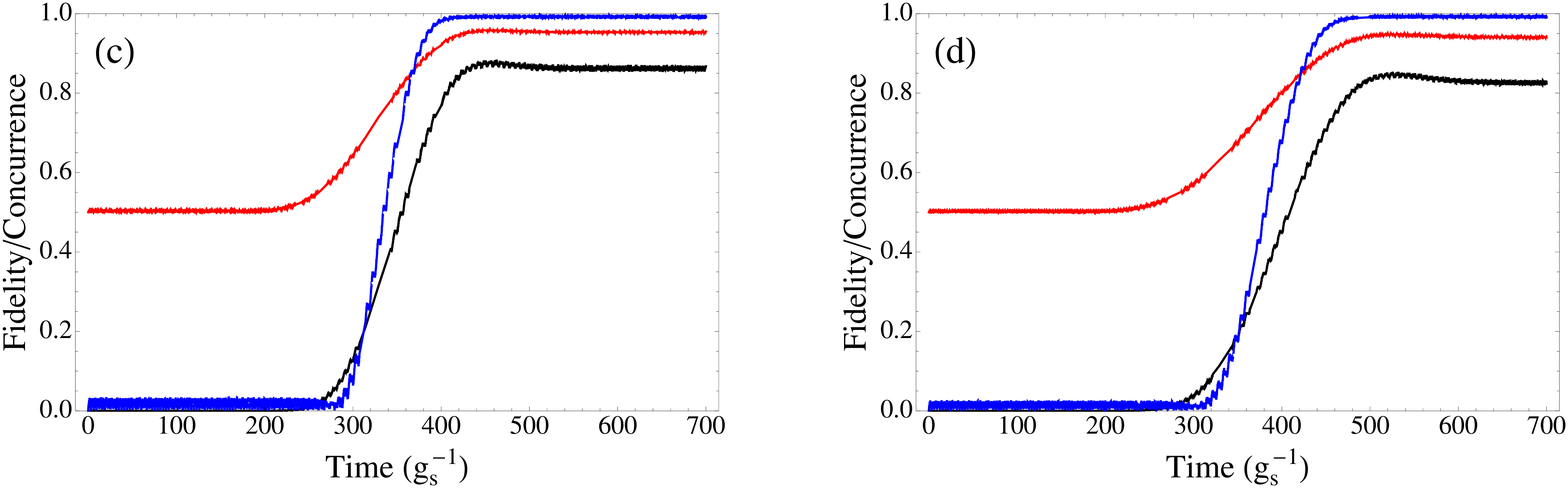}}
\caption{(Color online)  State transfer and entanglement generation mediated by a chain of 
$N$ spin qubits. 
The average fidelity $\bar{\cal F}$ (red),  the minimum fidelity ${\cal F}_{\min}$ (black), and  the concurrence (blue), 
are plotted as functions of time for $N=2$ (a,c) and $N=3$ (b,d). 
(a) Single-photon resonance ($\nu =1$),  and other 
parameters as in Fig. \ref{fig5}. (b) Single-photon resonance ($\nu =1$),  and other 
 parameters as in Fig. \ref{fig6}. (c)  Two-photon resonance ($\nu =2$), 
$z_0=3$, $g_{\rm s}\tau=41.42$ for $\Theta(T) =\pi/4$, and 
$g_{\rm s}\tau=82.85$ for $\Theta(T) =\pi/2$. Other 
parameters as in Fig. \ref{fig5}. (d) 
Two-photon resonance ($\nu =2$), 
$z_0=3$, $g_{\rm s}\tau=53.1$ for $\Theta(T) \simeq \pi/4$, and 
$g_{\rm s}\tau=106.2$ for $\Theta(T) \simeq\pi/2$. Other 
parameters as in Fig. \ref{fig6}.
The fidelities correspond to $\Theta(T) \simeq \pi/2$ and the concurrence for $\Theta(T) \simeq \pi/4$.
Time is in units of $g_s^{-1}$.  }
\label{fig7}
\end{figure*}

\subsection{Single-photon resonance}
\label{sec3a}
Consider first the case of terminal qubits whose energy difference 
matches the energy of a single photon i.e., we have 
$\omega_{\rm r} = \omega_{\rm s}+\omega$. The dynamics of the system for 
$N=2$ and $N=3$ spins in the channel are summarized in Figs. \ref{fig5} and \ref{fig6},   
where the populations of the sender $(|A_{\rm s}|^2)$, of the receiver $(|A_{\rm r}|^2)$, and of the channel $(1-|A_{\rm s}|^2-|A_{\rm r}|^2)$, are plotted as functions of time. In the case of continuous harmonic driving (i.e., for $f(t)=1$) the main part of the population oscillates back and forth between the terminal qubits [see black and blue curves in Figs. \ref{fig5}(a) and \ref{fig6}(a)], while the modes 
of the channel are scarcely populated. The dynamics are in very good agreement with the dynamics of the effective TLS as given by Eqs. (\ref{eom_tls:eqs}) 
[see dashed green curves]. The oscillations are slightly out of phase due to the ommission of 
rapidly rotating terms in the RWA. This is a well known effect, which is associated to 
the Bloch-Siegert shift when the neglected counter-rotating terms are small perturbations to the dynamics in the framework of RWA \cite{APbook,HuangPRA17}. 
This small phase difference is not expected to play a significantly role in 
the transfer of the state. Indeed, when the driving becomes pulsed, and the pulse duration is 
adjusted so that $\Theta(T) = \pi/2$, we find that about $90\%$ of the 
population has been transferred from the sender to the receiver at the end of the pulse 
[see Figs. \ref{fig5}(c) and \ref{fig6}(c)]. Moreover, when $\Theta(T) = \pi/4$ at the end of the 
pulse the 
population is distributed almost equally between the two terminal qubits. In either of the two cases, 
however, the final population of the receiver is below the ideal theoretically expected values, that is 
1 in the former case and 1/2 in the latter [corresponding to Eqs. (\ref{psi_pi2}) and (\ref{psi_pi4}), respectively]. There are mainly two reasons for these deviations.  

Firstly, a close inspection of Figs. \ref{fig5}(a) and  \ref{fig6}(a), reveals that when the driving is continuous the oscillations are not complete (i.e., the blue curves are always slightly below 1, and the black curves are slightly above 0). These findings suggest that for the parameters of the figures, the dynamics of our system are well approximated by a driven TLS where the photon energy does not  match precisely the difference of the shifted energies i.e.,  $\omega_{\rm s}-S_0\neq \omega_{\rm r} - S_1-\omega$. Indeed, if there was exact resonance, then the solutions of the amplitudes would be given by 
Eqs. (\ref{eom_tls:eqs}), and as depicted in Figs. \ref{fig5}(a) and \ref{fig6}(a), the oscillations would be  complete (see dashed green curve). 
 
Secondly, as depicted in the insets of Figs. \ref{fig5} and  \ref{fig6}, the populations exhibit oscillations after the end of the driving (i.e., for times $t>t_0+4\tau$ where $f(t)\simeq 0$). 
The larger oscillations stem from the fact that the terminal qubits are considered to be permanently coupled to the channel i.e., $g_{\rm s}$ and $g_{\rm r}$ do not vanish when the 
driving is over, and  one has to switch them off separately. This is part of the so-called read-out process and is present in almost every state transfer or entanglement generation protocol 
in spin-chains. The additional wiggles [mainly present in the inset of \ref{fig5}(b)] are due to the counter-rotating terms, which are present in the equations of motion (\ref{em1e_v1}). The fact that the dynamics do not freeze when the external driving of the sender is over, implies that the precise state of the receiver depends on the time 
$T\geq t_0+4\tau$, at which one decides to switch off the couplings $g_{\rm s}$ and $g_{\rm r}$. However, our 
numerical simulations suggest that the amplitude of these oscillations is very small, and thus the 
performance of our protocol is reliable irrespective of the chosen time $T$. 
To confirm this fact we have to quantify the performance of our protocol, in the cases of state transfer and entanglement generation.

As is the case for many other state-transfer protocols, the performance of our protocol depends on the qubit state to be transferred, and we are interested in reliable measures that are independent of the input state. To this end, two fidelity measures have 
been used widely in the literature, namely the  average-state fidelity 
and the minimum fidelity, which in the absence of disorder are given by 
\bea
\bar{\cal F} = \frac{1}{2}+\frac{|A_{\rm r}|^2}{6}+\frac{|A_{\rm r}|}{3}, 
\label{Fav:eq}
\eea 
and 
\bea
{\cal F}_{\min} = |A_{\rm r}|^2,  
\eea
respectively.
The former involves an average over all possible qubit states $\ket{\psi}_{\rm s}$, whereas the latter is   the lower bound on the fidelities that can be achieved for all possible $\ket{\psi}_{\rm s}$. 
As far as the entanglement generation is concerned, the performance of our scheme 
at any time $t>0$ can be quantified in terms of 
the concurrence ${\cal C}$ \cite{LPbook}, 
of the reduced density matrix for the terminal qubits at the particular time of interest. 
One readily obtains 
\bea
\rho (t)&=&
|A_{\rm s}|^2 \ket{1,0}\bra{1,0}
+|A_{\rm r}|^2 \ket{0,1}\bra{0,1}
+A_{\rm s} A_{\rm r}^\star \ket{1,0}\bra{0,1}\nonumber \\
&&+A_{\rm s}^\star A_{\rm r} \ket{0,1}\bra{1,0}
+(1-|A_{\rm s}|^2-|A_{\rm r}|^2 ) \ket{0,0}\bra{0,0},
\nonumber
\eea 
where for the sake of brevity we have set 
$\ket{1,0}\equiv \ket{1}_{\rm s}\otimes \ket{0}_{\rm r}$, and the concurrence is given by  
\bea
{\cal C}(t) = 2|A_{\rm s}(t)||A_{\rm r}(t)|. 
\label{Conc:eq}
\eea

All of the above measures are time dependent and their evolution for the parameters of Figs. 
\ref{fig5}  and \ref{fig6} is shown in Figs. \ref{fig7}(a,b). 
As expected, at the end of the pulse the fidelities and the concurrence exhibit small oscillations around some central values, which are considerably larger than the amplitude of the oscillations. 
Hence, the proposed protocol  achieves faithful state transfer and 
entanglement generation 
irrespective of the precise time (after the end of the driving), at which the user decides 
to switch off the couplings $g_{\rm s}$ and $g_{\rm r}$. Analogous results and dynamics have been obtained in the case of single-photon resonance 
for various combinations of $z_0$  and $N$.

\subsection{Multi-photon resonance}
\label{sec3b}

 The protocol also works when the energy difference of the terminal qubits is a multiple of the photon energy i.e., the resonance condition (\ref{resonance:eq}) is satisfied for some $\nu>1$. In this case we have multi-photon assisted processes, and the time evolution 
is analogous to the evolution depicted in Figs. \ref{fig5}, \ref{fig6} and Figs. \ref{fig7}(a,b).  
For instance, in Figs. \ref{fig7}(c,d), we plot the evolution of the aforementioned measures in the 
case of two-photon resonance, and for $z_0=3$. Clearly, the overall performance is slightly 
worse than for the single-photon case shown in Figs. \ref{fig7}(a,b), and the dynamics are 
about 1.6 times slower (compare the durations). 
By increasing $\kappa$, $|\omega_{\rm c}-\omega_{\rm s}|$ and $|\omega_{\rm c}-\omega_{\rm r}|$,  
the performance of the protocol can be improved, at the expense of even slower dynamics. 

Analogous results are expected for resonance with $\nu\geq 3$ photons, but as one can infer from the relative absolute values of the Bessel functions depicted in Fig. \ref{fig3},  in this case the dynamics will be considerably slower.

\section{Concluding Remarks}  
\label{sec4}

We have presented a protocol for photon-assisted state transfer 
and entanglement generation between two distant (spin)  qubits. The two terminal qubits 
have different energies, and are permanently coupled to an XX chain of $N$ far off-resonant spins, which is prepared 
in its ground (vacuum) state. The qubit of the sender is driven by an external pulsed harmonic field, and the process (state transfer or entanglement generation), is mediated by virtual excitation of the eigenmodes of the channel. 
The performance of our protocol has been investigated for various combinations of parameters, and the results presented here are a representative sample of our findings. We have shown that the present scheme allows for faithful state transfer and 
entanglement generation, when the  energy difference between the terminal qubits $\omega_{\rm r,s}$, is equal to or a multiple of the photon energy $\omega$. Hence, we have shown that it can perform reliably under conditions that go beyond the operational conditions of other related protocols i.e., when the energy difference is comparable to or larger than the couplings. For the sake of completeness as well as to account for possible scenarios where multi-photon processes may be preferable, the present formalism has been kept rather general, and our results apply to any number of photons. In principle, however, for a given energy difference $\omega_{\rm r,s}$ one can tune the amplitude of the driving $V_{\rm d}$ and the photon frequency $\omega$, so that the operation of our protocol is mediated by a single photon (i.e., $\omega_{\rm r,s}\simeq \omega$). 

A key feature of the present protocol, is that its operation relies solely on the external driving of the sender's qubit, while keeping all other couplings and energies constant. This may be an advantage for certain implementations of QIP, because one does not have to control a large number of qubits.  
Moreover,  at the end of the driving, the dynamics are frozen to a large extent, which facilitates the read-out process in a potential implementation. This is in contrast to other protocols (e.g., \cite{NikJPCM04,Christandl,weakRef1,Apollaro}), where the transferred (or generated)  state is localized 
at the receiver's site only for a small period of time, and the read-out process has to be fast.

During the evolution of the system the channel is  hardly populated; a  behaviour which is reminiscent of the stimulated Raman adiabatic passage (STIRAP) \cite{PetLamOC06,STIRAP}. 
There are, however, fundamental differences between our scheme and STIRAP. 
Firstly, by contrast to STIRAP, the operation of our scheme does not involve a dark state. Secondly, STIRAP requires the application of multiple pulses, whereas in our scheme only the 
qubit of the sender is driven. The scarce population of the channel is a common feature of all the adiabatic 
state-transfer schemes, including those of Refs. \cite{weakRef1,HarReuPle06,YaoPRL11}, and makes 
such protocols very robust against imperfections in the channel (including decoherence, dissipation 
and disorder). However, our protocol is fundamentally different from the ones in Refs. \cite{weakRef1,HarReuPle06,YaoPRL11}, because it involves  external driving of the sender's qubit; thereby  allowing for state transfer and entanglement generation between terminal qubits that differ in energy ($\omega_{\rm r,s} \neq 0$). 
It is only in the case of resonant terminal qubits (i.e., for $\omega_{\rm r,s} = 0$), where there 
is no need of driving. Indeed, in this case the resonance condition (\ref{resonance:eq}) is satisfied for $\nu=0$ photons, which means that state transfer and entanglement generation can be achieved without the assistance of photons.  
Hence, for resonant terminal qubits the driving  can be ignored, and one can readily confirm that the present scheme becomes practically equivalent to the adiabatic schemes of Refs. \cite{weakRef1, HarReuPle06}.

In general, our protocol becomes slower as we increase the number of spins in the channel $N$, because its operation relies on the virtual occupation of the channel. As we increase $N$, this condition requires larger 
energy differences between the terminal qubits and the channel, which result in a reduction of the 
Rabi frequency in the effective TLS. In practise, the characteristic time scale of decoherence is expected to determine the values of $N$ for which the protocol is attractive in the framework of a given physical implementation.  
For instance, consider a spin-based solid-state QIP architecture, where the qubit is the spin of an electron 
in GaAs quantum dot.  Typically, the exchange coupling in such a system is $g_{\rm s}\sim 0.2$meV, 
while the scale of the coherence (or the lifetime of the dot states themselves) is $T_2^\star\sim 20$ ns \cite{QD3,DasSarma03}. Hence, $T_2^\star\sim 
968 g_{\rm s}^{-1}$, whereas  the time scales required for state transfer and entanglement generation 
between five dots with our protocol are $T \sim 600 g_{\rm s}^{-1}$ in the case of single-photon resonance. 
The frequency of the driving falls in the microwave regime i.e.,  $\omega \sim g_{\rm s} \sim 48$GHz. 
These estimates are improved considerably, if one considers spin coherences of the order of $1-100\mu$s, as suggested in Ref. \cite{DasSarma03}.  For the time being, related experiments have been restricted to a small array consisting  of three tunnel-coupled quantum dots \cite{QD1}. These experiments have demonstrated the coherent oscillation of a 
charge between the two outermost dots, through the observation of Landau-Zener-St\"uckelberg interference. 
It is well known, however, that the transfer of the charge (excitation), does not necessarily imply the transfer of the associated spin state (which in addition to bit information carries phase information). Given the aforementioned decoherence times, the present results suggest that 
for the scheme of Ref. \cite{QD1}, the transfer of spin states, and the generation of entanglement are 
within reach of the current technology for arrays up to five dots.

The present results suggest that photons can provide new techniques for state transfer and entanglement generation in spin chains, which do not require any engineering, gate sequences or measurements. The protocol discussed in this work is certainly not unique, and there may be other 
photon-assisted protocols which perform better. 
There are still many questions to be addressed before one decides on the usefulness of 
photon assisted protocols, such as their performance in the presence of disorder, 
their applicability for various types of polarized and unpolarized spin chains \cite{YaoPRL11}, 
and  the exploitation of possible ways for speeding up the transfer of the state and the generation of entanglement. 
Finally, all of the results and discussion in this work pertain to parameters where the RWA is valid. 
If the RWA is not valid but the TLS is still a good approximation, then we have a TLS driven 
by terms involving many harmonics of the driving frequency (see also discussion in appendix \ref{app2}). In that case the 
optimal pulse duration may be obtained numerically by applying standard optimization techniques; where the 
interesting question is whether such optimization 
will lead to meaningful results, with the possibility of
faithful state transfer and entanglement generation.



\appendix

\section{Diagonalization of the Hamiltonian for the channel}
\label{app1}
The  Hamiltonian (\ref{HamHub}) can diagonalized by introducing the operators 
$
\hat{c}_{k} =\sum_{i=1}^{N} L_{k,i}\hat{a}_{i},  
$ \cite{weakRef1,ZwiAlvBenKurNJP14},  
with
\bea
L_{k,i}:= \sqrt{\frac{2}{N+1}} \sin\left ( \frac{ik\pi}{N+1}\right ).
\label{Lki:eq}
\eea
Hence, we have 
\begin{subequations}
\label{Hch:eq}
\bea
{\cal H}_{\rm c}  &=&  \sum_{k=1}^N \hbar\omega_k 
\hat{c}^{\dag}_k  \hat{c}_{k}  
\\
{{\widetilde{\cal V}}}  &=&  \sum_{k=1}^N \hbar  \left [ 
\tilde{g}_{{\rm s},k} \hat{a}_s^\dag \hat{c}_k + \tilde{g}_{{\rm r},k}  \hat{a}_{r}^\dag \hat{c}_k + {\rm H.c.} 
\right ]
\label{V2:eq}
\eea
\end{subequations}
with new coupling  $\tilde{g}_{{\rm s},k} := g_{{\rm s},k} e^{{\rm i}h(t)}$, 
$\tilde{g}_{{\rm r},k} := (-1)^{k-1}  g_{{\rm r},k} $, and 
\bea
g_{x,k} := g_{x} L_{k,1},\, \textrm{for}\,\, x\in\{{\rm r},{\rm s}\}.
\label{gsk:eq}
\eea
The eigenfrequencies are given by 
\bea
\omega_k = 2\kappa\cos\left ( \frac{k\pi}{N+1}\right )+\omega_{\rm c}.
\label{EigenEnergies:Eq}
\eea
Equation (\ref{EigenEnergies:Eq}) defines the so-called ``magnon" excitation energy. 
The interaction (\ref{V2:eq}) shows that 
the creation (annihilation) of a particle in the fermi state associated with the sender 
(or receiver) is accompanied by the destruction (creation) of magnons at different momenta, 
with different probabilities that are determined by the coupling coefficients. 

In the interaction picture \cite{LPbook}, the equations of motion for the amplitudes read
\begin{subequations}
\label{em1e}
\bea
{\rm i} \frac{{\rm d} A_{\rm s}}{{\rm d}t} &=& 
\sum_k \tilde{g}_{{\rm s},k} (t) 
e^{{\rm i}(\omega_{\rm s} - \omega_k) t}  A_k, 
\\
{\rm i} \frac{{\rm d} A_{\rm r}}{{\rm d}t} &=& 
\sum_{k=1}^N \tilde{g}_{{\rm r},k}  
e^{{\rm i}(\omega_{\rm r} - \omega_{k}) t}  A_k, 
\\
{\rm i} \frac{{\rm d} A_k}{{\rm d}t} &=& \tilde{g}_{{\rm s},k}^\star e^{-{\rm i}(\omega_{\rm s}-\omega_k) t} A_{\rm s} + \tilde{g}_{{\rm r},k} e^{-{\rm i}(\omega_{\rm r} - \omega_{k}) t}  A_{\rm r}.
\eea
\end{subequations} 
Applying the Jacobi-Anger expansion one obtains Eqs. (\ref{em1e_v2}), where 
\be
\tilde{g}_{{\rm s},k}^{(n)}(t):=g_{{\rm s},k}  {\cal J}_n ( z ). 
\label{gskn_tilde:eq}
\ee

\section{Reduction to an effective TLS}
\label{app2}
Formal integration of Eq. (\ref{em1e_v2:eq3}) 
yields 
\bea
 A_k(t) &=& -{\rm i}\sum_{n^\prime=-\infty}^{\infty} \int_{0}^t {\rm d} t^\prime 
 \tilde{g}_{{\rm s},k}^{(n^\prime)} (t^\prime) 
e^{{\rm i}(\omega_{k,{\rm s}}-n^\prime\omega) t^\prime} A_{\rm s}(t^\prime) 
\nonumber \\
&&-{\rm i}\tilde{g}_{{\rm r},k} 
\int_{0}^t {\rm d} t^\prime  e^{{\rm i}\omega_{k,{\rm r}} t^\prime}  A_{\rm r}(t^\prime).
\label{formal_int:eq}
\eea 
When conditions (\ref{All_assumptions}) are satisfied, the amplitudes $A_{\rm s}(t^\prime)$ and $A_{\rm r}(t^\prime)$ 
will not change much during the 
time the exponents in Eq. (\ref{formal_int:eq}) experience many oscillations.  
Assuming further that $f(t)$  changes in time sufficiently slow  [recall Eq. (\ref{ht2:eq})] 
so that 
\bea
\left | \frac{{\rm d}{\cal J}_{n^\prime}(t)}{{\rm d}t^\prime}\right | \ll |{\cal J}_{n^\prime}(t^\prime)|\cdot |\omega_{k,{\rm s}}-n^\prime\omega|
\label{Jslow:eq}
\eea
for all of the values of $n^\prime$ that dominate the dynamics and for all $t^\prime \in[0,t]$, 
we can evaluate $\tilde{g}_{{\rm s},k}^{(n^\prime)}$, $A_{\rm s}(t^\prime)$ and $A_{\rm r}(t^\prime)$  at time $t^\prime =t$. 

Hence, we have 
\bea
A_k(t)&\simeq & -\sum_{n^\prime=-\infty}^{\infty} \tilde{g}_{{\rm s},k}^{(n^\prime)}(t) 
\left 
[
\frac{e^{{\rm i}(\omega_{k,{\rm s}}-n^\prime\omega) t} -1}{\omega_{k,{\rm s}}-n^\prime\omega}
\right ] A_{\rm s}(t)  
\nonumber \\
&& -\tilde{g}_{{\rm r},k} 
 \left 
[
\frac{e^{{\rm i}\omega_{k,{\rm r}} t} -1}{\omega_{k,{\rm r}}}
\right ]
 A_{\rm r}(t).
\eea
This expression is substituted back in Eqs. (\ref{em1e_v2:eq1}) and (\ref{em1e_v2:eq2}). 
When conditions (\ref{All_assumptions}) are satisfied, one can neglect terms 
of the form $e^{\pm{\rm i}(\omega_{k,{\rm s}}- n\omega) t}$, and $e^{\pm {\rm i}\omega_{k,{\rm r}} t} $,  obtaining Eqs. (\ref{em1e_v3}) with 
\begin{subequations}
\bea
S_{0} &\simeq& \sum_k\sum_{n,n^\prime} 
\frac{\tilde{g}_{{\rm s},k}^{(n)}\tilde{g}_{{\rm s},k}^{(n^\prime)} }{\omega_{k,{\rm s}}-n^\prime\omega} 
e^{{\rm i}(n-n^\prime)\omega t} 
\\
S_{1} &\simeq& \sum_k \frac{(\tilde{g}_{{\rm r},k})^2}{\omega_{k,{\rm r}}} 
\\
\Omega_{0}&\simeq &
\sum_k\sum_{n} 
\frac{\tilde{g}_{{\rm s},k}^{(n)}\tilde{g}_{{\rm r},k} }{\omega_{k,{\rm r}}}
e^{-{\rm i}(\omega_{{\rm r},{\rm s}}-n\omega) t} 
\\
\Omega_{1} &\simeq& \sum_k\sum_{n^\prime} 
\frac{\tilde{g}_{{\rm s},k}^{(n^\prime)}\tilde{g}_{{\rm r},k} }{\omega_{k,{\rm s}}-n^\prime\omega}
e^{{\rm i}(\omega_{{\rm r},{\rm s}}-n^\prime\omega) t},
\eea
\end{subequations} 
and $(\tilde{g}_{{\rm r},k})^2 = (g_{{\rm r},k})^2$ in the expression for $S_1$.
The RWA has not been applied in the derivation of these expressions. 
As a result,  the summations 
extend from $-\infty$ to $\infty$, while some of them are time dependent. It is worth noting here that throughout this work we 
are interested in $0\leq z_0\leq 4$ and $|\min\{\omega_k\}-\omega_{\rm s}|\gg \omega$. Hence, we can ignore the divergence of $S_0$ and $\Omega_1$ for $n^\prime$ and $k$ such that $\omega_{k,{\rm s}} = n^\prime\omega$. This is because the sender and the eigenmodes are well separated in energy, and the denominators in $S_0$ and $\Omega_1$ may vanish only for large values of $n^\prime$ (see related discussion in Sec. \ref{sec2c}). At the same time, however, the enumerator is also very small because 
$|\tilde{g}_{{\rm s},k}^{(n^\prime)}|\propto |{\cal J}_{n^\prime}(\zeta)|\to 0$, as we increase $n^\prime$ (see Fig. \ref{fig3}). 
In practise, the eigenmodes of the channel are expected to acquire a finite lifetime, as a result of their coupling to the environment, and there will be no divergence. 

The above expressions for $S_{0(1)}$ and $\Omega_{0(1)}$ can be inserted in  
Eqs. (\ref{em1e_v3}), 
to obtain the dynamics of the effective TLS driven by a polychromatic field beyond RWA. 
However, when the $\nu-$photon resonance condition is satisfied [see Eq. (\ref{resonance:eq})], 
one can apply the RWA. The above expressions can be simplified farther and become time independent, thereby leading to the standard Rabi model with pulsed driving. 
We briefly describe here this reduction. 

A close inspection of the expression for $S_{0}(t)$ shows that for $n\neq n^\prime$ 
the various terms in the summations oscillate rapidly 
at frequencies $|n- n^\prime| \omega$, where $|n- n^\prime| \geq 1$. 
Similarly, for $n\neq \nu$, the terms in the summations of $\Omega_{0(1)}(t)$ oscillate 
rapidly at frequencies $|n-\nu|\omega$, where $|n-\nu|\geq 1$. As long as we are interested in the dynamics of the TLS on time scales $T\gg \omega^{-1}$, we can approximate $S_{0}(t)$ and $\Omega_{0(1)}(t)$ 
in Eqs. (\ref{em1e_v3}) by their time-average values. The terms associated with the resonance 
are expected to dominate the time-average, whereas all of the remaining fast oscillating terms 
are expected to cancel out thereby yielding a negligible contribution. 
More precisely, assuming without loss of generality that   $\omega_{{\rm r},{\rm s}}>0$ and $\omega_{k,{\rm s}}>0$, the terms that will survive pertain to 
$n=n^\prime = \nu$. Hence, in the denominators we will have $\omega_{k,{\rm s}}-n^\prime\omega\simeq \omega_{k,{\rm r}}$, while all of the rapidly oscillating terms are neglected. Hence, we obtain 
\bea
S_{0} &\simeq& \sum_k \frac{(g_{{\rm s},k})^2}{\omega_{k,{\rm r}}}, 
\\
\Omega_{0}^{(\nu)}&=&\Omega_{1}^{(\nu)} \simeq  \sum_k
\frac{\tilde{g}_{{\rm s},k}^{(\nu)}\tilde{g}_{{\rm r},k} }{\omega_{k,{\rm r}}},
\eea
with $\tilde{g}_{{\rm s},k}^{(\nu)}$ given by Eq. (\ref{gskn_tilde:eq}).
In the derivation of these expressions we have assumed that Eq. (\ref{resonance:eq}) is satisfied. 
In general, the RWA is expected to be valid when $|\nu\omega-\omega_{\rm r,s}|\ll \omega,\quad \Omega_{0(1)}^{(\nu)}\ll\omega $.  

In closing, it is worth noting  that $S_0$ is not expected to coincide exactly with $S_1$, even when 
$g_{\rm s} = g_{\rm r}$, because $S_0$ involves additional approximations that have not been 
applied in $S_1$. Hence, even if the photon energy is such that a multiple of it matches exactly the 
energies of the undriven system [i.e., condition (\ref{resonance:eq}) is satisfied exactly for some $\nu$], 
one should expect a detuning between the shifted levels of the driven TLS 
and the $\nu-$photon energy $\nu\omega$, which is given by $\Delta = S_0-S_1$.  
However, one is close to resonance when $\Omega_{0(1)}^{(\nu)}\gg |\Delta |$. 
Our simulations suggest that the detuning depends weakly on the number of qubits in the chain $N$. On the other hand, the Rabi frequencies $\Omega_0^{(\nu)}$ and $\Omega_1^{(\nu)}$ are always very close to each other, and they decrease with increasing $N$, due to the presence of the term $(-1)^{k-1}$ in $\tilde{g}_{{\rm r},k}$, which causes alternating signs in the Rabi frequencies.

\end{document}